# LOCAL NONLINEAR APPROXIMATIONS TO THE GROWTH OF COSMIC STRUCTURES

Paul J. Mancinelli[1] and Amos Yahil[1,2]

## ABSTRACT

Local nonlinear approximations to the growth of cosmic perturbations are developed, resulting in relations, at a given epoch, between the peculiar velocity and gravity fields and their gradients. Only the equation of motion is approximated, while mass conservation and the computation of the gravitational field are treated exactly. The second-order relation is derived for arbitrary geometry and cosmological parameters. Solutions are developed to fourth order for laminar spherical perturbations in an Einstein-de Sitter universe, but the gain in accuracy for higher orders is modest. All orders become comparable when the peculiar kinetic energy per unit mass equals the peculiar potential, typically at relative density perturbations, $\delta \sim 4$. The general second-order relation, while implicit, is simple to solve. $N$-body simulations show that it provides moderate gains in accuracy over other local approximations. It can therefore be easily applied in the comparison of large-scale structures and velocities in the quasi-linear regime, $\delta \sim 1 - 4$, as well as in the reconstruction of the primordial perturbations from which they grew.

*Subject headings:* cosmology: theory — gravitation — instabilities — large-scale structure of universe

## 1. INTRODUCTION

The need for practical nonlinear approximations to the growth of cosmic perturbations is widely recognized. Initial-value problems, in which positions and velocities are given at some initial epoch and the system is evolved forward in time, are now routinely computed into the nonlinear regime using $N$-body codes. Shortcut approximations to initial-value problems therefore find practical application only when the resolution required exceeds that available with $N$-body codes, or when a large statistical sample is needed and the computational effort of many $N$-body simulations is prohibitive.

---

[1] Astronomy Program, State University of New York, Stony Brook, NY 11794–2100, U.S.A.

[2] Yukawa Institute for Theoretical Physics, Kyoto University, Kyoto 606, Japan

But initial-value schemes, including $N$-body codes, are unable to solve problems with mixed boundary conditions, for which some positions and velocities are given at one epoch and some at another epoch. Such problems arise in the comparison of large-scale density and velocity fields and need to be solved in the quasi-linear (mildly nonlinear) regime in order to separate the effects of cosmology from those of biased galaxy formation (Dekel *et al.* 1993). Also, $N$-body codes are unstable when integrating cosmological systems backward in time to high redshift, and nonlinear approximations are therefore needed for "time machines" which seek to reconstruct the initial density and velocity fields from which the current large-scale structure grew (Nusser & Dekel 1992; Gramann 1993a).

It is instructive to examine where the nonlinear effects first manifest themselves in a growing perturbation. This can be done by substituting the linear solution into the omitted nonlinear terms of the equations which govern gravitational instability and seeing when they become comparable to the other terms. Not only is the linear approximation invalid at that point, but it is likely that any perturbative expansion will contain many, possibly an infinite number, of comparable terms, i.e., this sets the radius of convergence of the expansion. Of the three equations for Newtonian pressureless gravitational instability, the Poisson equation is always linear, the nonlinear term in the continuity equation becomes comparable to the linear one when the relative density perturbation, $\delta \equiv \delta\rho/\rho \sim 1$, and the Euler equation of motion reaches this breakpoint when $(\mathbf{v} \cdot \nabla)\mathbf{v} \sim \mathbf{g}$, where $\mathbf{v}$ and $\mathbf{g}$ are the peculiar velocity and gravity, respectively. Since the Kelvin circulation theorem ensures that the flow is irrotational, as long as it remains laminar, the velocity gradient term can be rewritten as $\nabla(\frac{1}{2}v^2)$. The linearized Euler equation is therefore expected to fail when the peculiar kinetic energy per unit mass equals the peculiar gravitational potential. $N$-body simulations, as well as simple analytical models, e.g., a spherical top-hat perturbation, show that this typically occurs when $\delta \sim 4$.

In order to extend an approximation into the quasi-linear regime, $\delta \sim 1 - 4$, it is necessary to solve the continuity equation exactly at the onset of nonlinear effects when $\delta \sim 1$, or to enforce mass conservation in some other way. One way to to do this is to follow the trajectories of all mass points, i.e., a Lagrangian approach. Zel'dovich (1970) first pointed out that in the linear regime, the displacement vector can be separated into a product of a universal time function and a function of initial position

$$\mathbf{x}(t) = \mathbf{q} + D(t)\mathbf{C}(\mathbf{q}) \quad , \tag{1}$$

where $\mathbf{x}$ and $\mathbf{q}$ are the current and primordial comoving coordinates, and $D(t)$ is the linear growth factor, i.e., $\delta \propto D$. Zel'dovich further suggested that Eq. (1) might also be a good nonlinear approximation. His argument was that the displacement field can be well behaved and lend itself to perturbative approximations, while the density fluctuates strongly at points of convergent flow. Therefore, Eulerian perturbation theories, which expand the equations in powers of density and velocity, are likely to have a much more limited range of validity, while Lagrangian expansions of the displacement vector might be valid for $\delta \gtrsim 1$.

Numerical simulations have generally borne out his hypothesis (e.g., Melott, Pellman, & Shandarin 1994, and references therein).

Recently there has been a resurgent interest in higher-order Lagrangian perturbation theories in the hope of improving Eq. (1). There are three broad categories of such methods: formal perturbation theories, action methods, and "local" approximations.

A formal theory consists of expanding the continuity, Euler, and Poisson equations in powers of a perturbation parameter $\epsilon$ and equating coefficients power by power (Bouchet *et al.* 1994, hereafter BCHJ, and references therein; Buchert 1994, and references therein; Catelan 1994). The first order recovers the Zel'dovich approximation, and higher orders provide corrections. These methods are generally suitable only for initial-value problems.

Peebles (1989) proposed to solve mixed-boundary-condition problems using Hamilton's principle: expand the displacement vector for each mass point as a sum of functions of time with unknown coefficients and determine the coefficients by seeking orbits which render the action stationary. While the expansion functions are arbitrary in principle, a judicious choice can improve convergence dramatically and help disentangle the multiple solutions of mixed-boundary-condition problems. Giavalisco *et al.* (1993, hereafter GMMY) pointed out that linear theory requires the first-order function to be $D(t)$ and suggested that the higher-order terms be simply powers of $D$. They obtained excellent fits to the nonlinear spherical case, the most difficult for this scheme, using second, third and fourth order expansions.

Under the category of local approximations we include all schemes which relate **v**, **g**, and their spatial derivatives at a given epoch. Several such methods have been proposed recently, some based on dynamical arguments, some purely phenomenological (Nusser *et al.* 1991, hereafter NDBB; Bernardeau 1992; Gramann 1993b). Mancinelli *et al.* (1994) checked their accuracy using $N$-body simulations and found them to be comparable, with a slight advantage to the NDBB approximations.

We propose a new, local, higher-order Lagrangian perturbation theory, §2, which adheres to the spirit of the original Zel'dovich approximation in that mass conservation and the Poisson equation are treated exactly, and only the Euler equation is approximated. We first show that the peculiar velocity and gravity fields can be expanded in a simple way, so that they are related to each other term by term. The exact continuity and Poisson equations are then used to couple the different terms, giving an implicit relation between peculiar gravity and velocity. We derive the general second-order relation between the two fields.

While this new approximation can in principle be extended to any order, in practice, we find the added algebraic complexity outweighs the accuracy gained by expanding to higher orders. In §3 we investigate the usefulness of the third and fourth orders for spherical perturbations ($\Omega = 1, \lambda = 0$) and find the increase in accuracy to be modest. In fact, we confirm the heuristic argument given above that all orders become comparable when the peculiar kinetic energy per unit mass equals the peculiar gravitational potential, which occurs at $\delta \sim 4$, thus limiting the usefulness of expanding to higher orders.

In §4 we check the accuracy of our general second-order approximation by comparing it with the same $N$-body simulations used by Mancinelli *et al.* (1994). We find it to be a minor improvement over the NDBB approximations. We summarize and discuss other applications in §5.

## 2. THE APPROXIMATION

In a pressureless universe, the equation of motion of a mass element in the non-inertial comoving frame is
$$\frac{d}{dt}(a\mathbf{v}) = a\mathbf{g} \quad , \tag{2}$$
where $\mathbf{v}$ is the peculiar velocity relative to the comoving frame and $\mathbf{g}$ is the peculiar acceleration which is purely due to density perturbations and does not include the effects of any smooth component of the density or a cosmological constant. Note that the smooth density background affects the equation of motion only through the time dependence of the expansion parameter $a$.

It is useful to change variables from $t$ to $D$ and to rescale $\delta$, $\mathbf{v}$ and $\mathbf{g}$ such that
$$\Delta \equiv \frac{\delta}{D} \quad , \tag{3}$$

$$\mathbf{V} \equiv \frac{d\mathbf{x}}{dD} = \frac{\mathbf{v}}{a\dot{D}} \quad , \tag{4}$$

$$\mathbf{G} \equiv \frac{a}{\Omega \dot{a}^2 D}\mathbf{g} \quad . \tag{5}$$

Rewriting Eq. (2) in terms of the rescaled velocity and gravity, and using the differential equation for $D$,
$$\frac{d}{dt}(a^2 \dot{D}) = \frac{3}{2}\Omega \dot{a}^2 D \quad , \tag{6}$$
yields
$$\frac{d\mathbf{V}}{dD} = \frac{\Omega}{f(\Omega,\lambda)^2 D}\left(\mathbf{G} - \frac{3}{2}\mathbf{V}\right) \quad , \tag{7}$$
where
$$f(\Omega,\lambda) = \frac{d\ln D}{d\ln a} \approx \Omega^{0.6} + \frac{\lambda}{70}\left(1 + \frac{\Omega}{2}\right) \tag{8}$$
is the logarithmic growth factor and $\lambda$ is the dimensionless cosmological constant (Peebles 1976; Lahav *et al.* 1991).

Eq. (7), which is exact, is our starting point. For growing perturbations, all variables are analytic in $D$ at $D = 0$. It follows that the expression in brackets on the r.h.s. of Eq. (7) must vanish at least as $D$, which gives

$$\mathbf{V}_p - \frac{2}{3}\mathbf{G}_p = 0 \quad , \tag{9}$$

where the subscript $p$ denotes the primordial value. This is the linear approximation.

In the original Zel'dovich approximation $\mathbf{V}$ is constant, so Eq. (9) must hold at all times. We depart from this limit by considering higher-order terms. Note that for $\Omega \neq 1$, the factor $f^2/\Omega \approx \Omega^{0.2}$ in Eq. (7), although varying slowly as a function of $\Omega$, does introduce a time dependence due to the changing curvature of the universe. This dependence can be eliminated by rewriting Eq. (7) in terms of another variable, $E$, given by

$$\frac{dE}{dD} = \frac{\Omega E}{f^2 D} \quad , \qquad \lim_{D \to 0} \frac{E}{D} = 1 \quad , \tag{10}$$

which removes the $\Omega$ dependence from Eq. (7) to give

$$\frac{d\mathbf{V}}{dE} = \frac{1}{E}\left(\mathbf{G} - \frac{3}{2}\mathbf{V}\right) \quad . \tag{11}$$

Since $dE/dD = 1$ at $D = 0$, it follows that $\mathbf{V}$ and $\mathbf{G}$ are also analytic in $E$ and can therefore be Taylor expanded in $E$. The expansion coefficients of the two fields around $E = 0$ are simply related order by order,

$$\mathbf{G}_p^{(n)} = (n + 3/2)\mathbf{V}_p^{(n)} \quad , \tag{12}$$

where $(n)$ denotes the $n$'th derivative with respect to $E$. The coupling between the different orders is obtained by applying the continuity and Poisson equations. This provides an initial-value perturbation expansion of the fields in terms of one independent term, set by the initial conditions.

For mixed-boundary-condition problems one would like to expand the fields around a finite $E$. This is most easily achieved by integrating Eq. (11) to give

$$\mathbf{V}(E) = E^{-3/2}\int_0^E dE' E'^{1/2}\mathbf{G}(E') \quad , \tag{13}$$

expanding $\mathbf{G}(E')$ in a Taylor series around $E$, and explicitly evaluating the integrals in the sum to obtain

$$\mathbf{V}(E) = \sum_{n=0}^{\infty}(-)^n\frac{\Gamma(3/2)}{\Gamma(n+5/2)}E^n\mathbf{G}^{(n)}(E) \quad . \tag{14}$$

For the second-order approximation we ignore second and higher derivatives of $\mathbf{G}$. We can then use Eq. (10) to restore $D$ as our variable, obtaining

$$\mathbf{V} = \frac{2}{3}\mathbf{G} - \frac{4f^2}{15\Omega}D\frac{d\mathbf{G}}{dD} \quad . \tag{15}$$

Eq. (15), or the more general Eq. (14), are not yet in useful form, since we do not usually know the time derivatives of the gravitational field. We can, however, use the Poisson and continuity equations to convert the time derivatives to spatial gradients, thus obtaining an implicit *local* relation between $\mathbf{G}$, $\mathbf{V}$, and their gradients. To do this we take the divergence of Eq. (15), and, since the divergence operator does not commute with the Lagrangian (convective) time derivative $d/dD$, convert it to Eulerian derivatives to yield

$$\nabla \cdot \frac{d\mathbf{G}}{dD} = \frac{\partial}{\partial D} \nabla \cdot \mathbf{G} + \nabla \cdot (\mathbf{V} \cdot \nabla)\mathbf{G} \qquad . \tag{16}$$

Applying the Poisson equation

$$\nabla \cdot \mathbf{G} = -\frac{3}{2}\Delta \quad , \tag{17}$$

and the continuity equation

$$D\frac{\partial \Delta}{\partial D} = -\Delta - \nabla \cdot (1 + D\Delta)\mathbf{V} \quad , \tag{18}$$

and taking advantage of the irrotational nature of the gravitational field, i.e., $\nabla \times \mathbf{G} = 0$, we obtain, after some algebra,

$$D\nabla \cdot \frac{d\mathbf{G}}{dD} = -\nabla \cdot \mathbf{G} + \frac{3}{2}\nabla \cdot \mathbf{V} + D\left(G_{i,j} - G_{k,k}\delta_{ij}\right)V_{i,j} \quad , \tag{19}$$

where a subscript preceded by a comma indicates an Eulerian derivative with respect to that component, $\delta_{ij}$ is the Kronecker $\delta$ symbol, and the tensor summation convention is implied.

Substituting Eq. (19) into the divergence of Eq. (15) and rearranging, we obtain our final result:

$$\nabla \cdot \mathbf{V} - \frac{2}{3}\nabla \cdot \mathbf{G} = -\alpha D\left(G_{i,j} - G_{k,k}\delta_{ij}\right)V_{i,j} \quad , \tag{20}$$

where the nonlinear coefficient

$$\alpha = \frac{4}{6 + 15\Omega/f(\Omega,\lambda)^2} \quad . \tag{21}$$

Since $f(\Omega, \lambda)$, Eq. (8), is exceedingly insensitive to $\lambda$, the nonlinear coefficient $\alpha$ is as well. As a function of $\Omega$, $\alpha$ ranges between 0.13 and 0.21 for $0.1 < \Omega < 2$, with a typical value of $4/21 \approx 0.19$ for $\Omega = 1$. The nonlinear correction is therefore also insensitive to $\Omega$, a point often noted in previous work.

Gramann (1993ab) took a similar approach to ours, but approximated the continuity equation, thereby obtaining explicit expressions for $\mathbf{V}$ in terms of $\mathbf{G}$ or vice-versa. We prefer to achieve higher accuracy by conserving mass exactly and leave the relation between $\mathbf{V}$ and $\mathbf{G}$ in its implicit form. Also, the weak dependence of the nonlinear coefficient on $\Omega$ is different in the two approximations, since we expand $\mathbf{V}$ and $\mathbf{G}$ in $E$, while Gramann expands in $D$.

The nonlinear term on the r.h.s. of Eq. (20) can be split into a scalar term, which is related to the density via the Poisson equation, and a traceless, symmetric shear term, giving

$$(1 + \alpha D \Delta) \nabla \cdot \mathbf{V} + \alpha D \left( G_{i,j} - \frac{1}{3} G_{k,k} \delta_{ij} \right) V_{i,j} = -\Delta \quad . \tag{22}$$

Without the shear term Eq. (22) is identical to the phenomenological expression of NDBB, their Eq. (38), and provides the first dynamical justification for it. Their nonlinear coefficient, estimated experimentally from $N$-body simulations to be 0.18, is close to the one derived here. The importance of the shear term, which they did not include, can be seen by considering laminar planar perturbations. In this case $\mathbf{G}$ is time independent and linear theory holds exactly (Doroshkevich, Ryabenki, & Shandarin 1973). Eq. (14) shows that this is satisfied to all orders, and, indeed, the nonlinear term on the r.h.s. of Eq. (20) vanishes. But the scalar and shear terms in Eq. (22) separately are nonzero, and the omission of either one introduces an error. A second problem with the NDBB approximation is that the average of Eq. (22) over a large volume does not vanish without the inclusion of the shear term. In fact, in order to correct for this effect in applications of the NDBB approximation, a constant offset has usually been applied to $\Delta$.

Given a density field, and hence $\mathbf{G}$, Eq. (20) is a linear equation for $\mathbf{V}$ with non-constant coefficients. Conversely, given $\mathbf{V}$ it is a linear equation for $\mathbf{G}$, also with non-constant coefficients. It is therefore suitable for problems with mixed boundary conditions, in which one solves for one field in terms of the other. To be more precise, only the irrotational part of the velocity field can be determined in this way. For laminar flow, however, the Kelvin circulation theorem guarantees that the flow is, in fact, irrotational, and the problem can be restated in terms of velocity and gravitational potentials.

If it were not for the shear term, the solution of Eq. (22) would be trivial, since the non-constant part factors out and can be treated as a source term; in this case, one is solving a modified Poisson equation. This simple procedure is not possible in the presence of the shear term. There are two ways to overcome this difficulty. One can solve the complete $N \times N$ linear problem, where $N$ is the number of grid points. This is not as prohibitive as might be thought, since the coefficient matrix is very sparse and symmetric. In fact, the number of operations is $O(N)$, comparable to FFT computations. We are in the process of constructing an efficient code to solve such linear systems. A poor man's alternative, which seems to work well (§4), is to treat the shear as a source term and to solve for it by iteration.

## 3. HIGHER-ORDER SPHERICAL APPROXIMATIONS

Higher-order approximations can be derived in an analogous manner to the second-order one obtained in §2, but they quickly become very cumbersome. Here we investigate the possible benefit of such higher-order approximations by considering the case of a laminar spherical perturbation for $\Omega = 1$ and $\lambda = 0$.

In this case
$$G = -\frac{1}{2}\Delta \mathbf{x} \quad , \qquad (23)$$
and the internal mass distribution inside a shell is irrelevant. Therefore, without loss of generality, we can take a top-hat perturbation with constant $\Delta$ as a function of radius. After expanding Eq. (14) up to third and fourth orders and computing the necessary derivatives of $\mathbf{G}$, we obtain the implicit relations

$$\tfrac{8}{189}(1 + D\Delta)D(\nabla \cdot \mathbf{V})^2 + \left(1 + \tfrac{8}{27}D\Delta\right)(\nabla \cdot \mathbf{V}) + \left(1 + \tfrac{4}{63}D\Delta\right)\Delta = 0 \quad , \qquad (24)$$

and

$$\begin{aligned}\tfrac{64}{6237}(1 + D\Delta)D^2(\nabla \cdot \mathbf{V})^3 + \tfrac{64}{693}(1 + D\Delta)D(\nabla \cdot \mathbf{V})^2 \\ + \left(1 + \tfrac{284}{693}D\Delta + \tfrac{88}{2079}D^2\Delta^2\right)(\nabla \cdot \mathbf{V}) + \left(1 + \tfrac{8}{63}D\Delta\right)\Delta = 0 \quad ,\end{aligned} \qquad (25)$$

respectively.

The quadratic Eq. (24) has a real solution for $\nabla \cdot \mathbf{V}$ for $\Delta \lesssim 4.55$ (very close to the turnaround density contrast), while the cubic Eq. (25) always has a real solution. These solutions, as well as the second-order approximation, are plotted in the bottom panel of Fig. 1 as $-\nabla \cdot \mathbf{v}/\delta$ ($\equiv 1$ in the linear approximation) versus $1 + \delta$. For comparison we also show the exact solution and the second-order L2 approximation of BCHJ. We see that the velocity approximation is improved somewhat with higher orders for $\delta \lesssim 4$, but convergence is very slow for higher densities, and all approximations become comparable at $\delta \sim 4$. This confirms the argument given in §1 that the radius of convergence of the expansion is reached at the point where the peculiar kinetic energy per unit mass equals the peculiar gravitational energy, at $\delta \sim 4$. The approximation of BCHJ is comparable to our second-order approximation for $\delta > 0$, but is poorer for $\delta < 0$.

By contrast, we show in the upper panel of Fig. 1 successive approximations of the action method (GMMY), for which Hamilton's principle guarantees convergence to the exact solution. These approximations are indeed seen to converge to the exact solution more rapidly and over the entire range of $\delta$. Another way to understand this convergence is to note that Eq. (14) approximates the integral in Eq. (13) by a quadrature based on end-values of $\mathbf{G}$ and its time derivatives, which are converted to spatial derivatives. The action integral, on the other hand, is influenced by the entire integration interval, making it effectively a time-centered quadrature, which improves convergence.

We conclude that for applications for which the second-order approximation of §2 is inadequate, higher orders of Eq. (14) are unlikely to provide a significant improvement, and an action method is preferable.

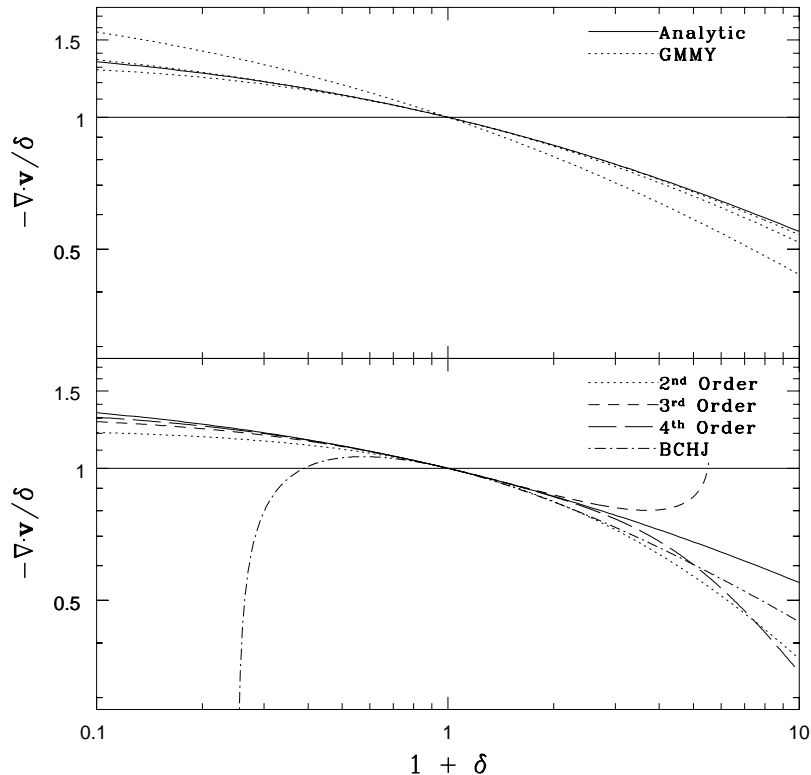

Fig. 1.— The quantity $-\nabla \cdot \mathbf{v}/\delta$ ($\equiv 1$ for linear theory) as a function of density for a spherical perturbation in an Einstein-de Sitter universe. The upper panel shows the action method of GMMY up to third order (dotted) rapidly converging to the analytic solution (solid). The lower panel compares the same analytic solution with the second (dotted), third (short dashes) and fourth (long dashes) order solutions presented in this paper and the second-order L2 approximation of BCHJ.

## 4. N-BODY SIMULATIONS

In a previous study (Mancinelli et al. 1994) we compared various local approximations, using three independent $N$-body simulations of a CDM universe with $\Omega = 1$, $\lambda = 0$, and $h = 0.5$ ($H_0 = 100\ h$ km s$^{-1}$ Mpc$^{-1}$). We found them to be comparable, with a small advantage to the NDBB approximations, and therefore compare our new second-order approximation against NDBB, using the same simulations run by Mancinelli et al. . These simulations were identical except for the choice of the seed of the random number generator used to create the initial conditions, and were run on a $128^3$ grid with comoving spacing of 200 km s$^{-1}$. The initial perturbations were normalized in the standard way to unit variance in a sphere of radius 800 km s$^{-1}$, if extrapolated linearly to the present epoch. They were then integrated forward using a particle-mesh (PM) code. The resultant, fully nonlinear density and velocity fields at the present epoch were computed at the grid points using a cloud-in-cell method, followed by Gaussian smoothing; we refer to them as the "exact"

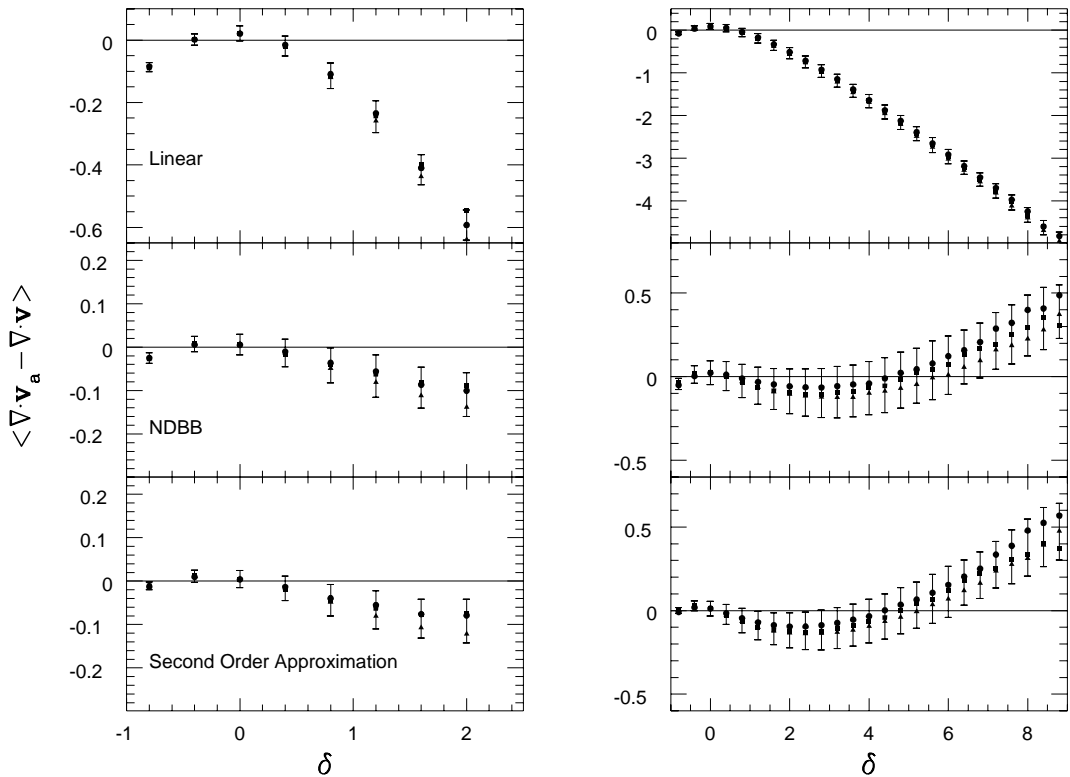

Fig. 2.— The average differences between the linear, NDBB, and our second-order approximations and the exact divergence of the velocity field obtained from $N$-body simulations of a CDM universe, Gaussian smoothed with a radius of 1000 km s$^{-1}$ (left-hand panels) and 500 km s$^{-1}$ (right-hand panels), with grid spacing of 200 km s$^{-1}$ for both smoothing radii. The points are the *mean* values obtained in each density bin in three identical simulations, differing only in the seed of the random number generator which created the initial conditions. The error bars are a measure of the dispersion in the approximations of *individual* data (grid) points. See the text for details.

fields. Eq. (20) was solved iteratively, with previous iterations used in the nonlinear term.

The second-order approximation of §2 for the divergence of the velocity field, given a density field, are evaluated in Fig. 2, where the left-hand panels are for a Gaussian smoothing radius of 1000 km s$^{-1}$ and the right, 500 km s$^{-1}$. For comparison we also show the linear approximation, but note the different scale. Here we plot the differences between the approximate and exact divergences, $(\nabla \cdot \mathbf{v}_a - \nabla \cdot \mathbf{v})$. The points are the *mean* differences for each of the three simulations and show the systematic error of each approximation as a function of the (exact) density. The error bars in the figures are a measure of the dispersion in the approximations of *individual* data (grid) points. They were measured by taking the *r.m.s.* cross differences of $(\nabla \cdot \mathbf{v}_a - \nabla \cdot \mathbf{v})$ between points in independent simulations. Grid points in the same simulation may not be used for this purpose because they may fall within each other's smoothing radius, causing an underestimate of the dispersion. For the same reason, the standard deviations of the means are not equal to the standard deviations divided by $\sqrt{N}$, nor are the means of different bins independent. It is therefore better to estimate the uncertainties in the means from the scatter between different simulations; we have chosen simply to plot the means of all three simulations. Fig. 3 shows the analogous comparison for velocity fields approximated from densities. The quantity for comparison here is $|\mathbf{v}_a - \mathbf{v}|^2$, except that we plot its square root in order to express it in km s$^{-1}$.

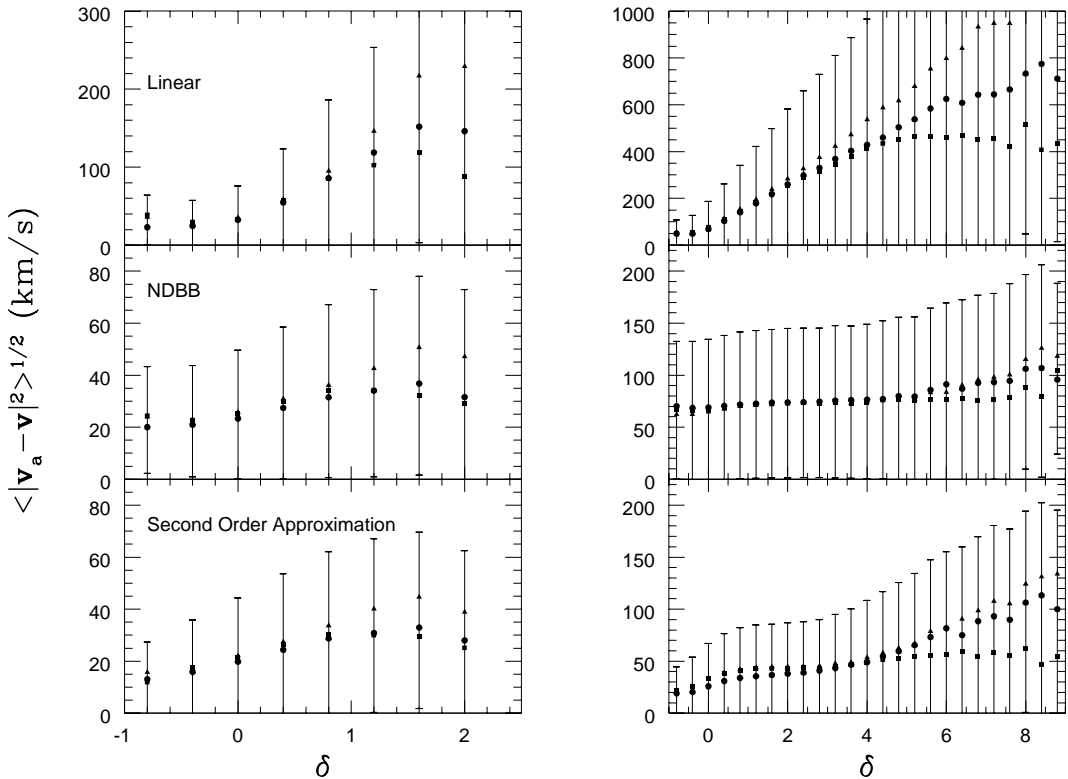

Fig. 3.— Same as Fig. 2, but the quantity plotted here is the average of the *r.m.s.* of the difference of the approximate and the exact velocity fields.

The first item to note in Fig. 2 is that both nonlinear approximations are a vast improvement over linear theory, which breaks down quite rapidly for $\delta > 1$. For $\nabla \cdot \mathbf{v}$ they are nearly identical in the range $0 < \delta < 10$. For $\delta < 0$, our approximation more accurately reproduces $\nabla \cdot \mathbf{v}$, particularly in the last bin. This can be attributed to the shear term, which the NDBB approximation lacks.

Fig. 3 shows a similar improvement over linear theory for both approximations. For smoothing of 1000 km s$^{-1}$ (left-hand panels), the velocity field predicted by our approximation is a slight improvement over NDBB for $\delta < 0$, with $\langle |\mathbf{v}_a - \mathbf{v}|^2 \rangle^{1/2} \lesssim 20$ km s$^{-1}$. For positive $\delta$, the two approximations predict the velocity field to roughly the same accuracy. For smoothing of 500 km s$^{-1}$ (right-hand panels), our approximation predicts the velocity to within 50 km s$^{-1}$ in the range $-1 < \delta < 5$, whereas the NDBB approximation shows errors of $\sim 70$ km s$^{-1}$ over this range.

We also computed errors for the inverse problem: predict the density field given the velocity field. The errors, $\delta_a - \delta$, are quite similar to the results of Fig. 2, with only a slight increase in the scatter at the high $\delta$ end, and are not reproduced here.

## 5. SUMMARY AND DISCUSSION

We have laid the groundwork for higher-order local approximations of the growth of gravitational instabilities, in which $\mathbf{v}$, $\mathbf{g}$, and their gradients are related at a given epoch. Our method approximates the equation of motion, but computes gravity and enforces mass conservation exactly. We have derived the second-order solution for an arbitrary geometry and cosmological model and find it to be similar to the phenomenological approximation of NDBB and the second-order solution of Gramann (1993b). It differs from the former by the addition of a shear term, and from the latter by requiring exact mass conservation. $N$-body simulations show that it provides a modest improvement in accuracy.

Unlike $N$-body codes and initial-value approximations, both local approximations and action methods are suitable for mixed-boundary-condition problems. The comparison of large-scale structures and velocities at the present epoch is a direct application of a local approximation such as Eq. (22). The reconstruction of the primordial perturbation fields is also straightforward. For example, the second-order approximation, Eq. (20), can be integrated to give $(\mathbf{G} - 3\mathbf{V}/2)/D$, which is finite as $D \to 0$. This can then be used as a source term in Eq. (7) to integrate $\mathbf{V}$ back in time from the present epoch, which is analogous to the method of Nusser and Dekel (1992).

We have also computed solutions up to fourth order for spherical perturbations in an Einstein-de Sitter universe and find that the added algebraic complexity outweighs the increase in accuracy. All orders become comparable when the peculiar kinetic energy per unit mass equals the peculiar potential, typically at $\delta \sim 4$. This is in stark contrast to the action method of GMMY, which quickly converges with the addition of higher orders for all density contrasts.

The advantage of the local approximation proposed here is its simplicity: a direct relation between $\mathbf{v}$, $\mathbf{g}$, and their gradients. Although it is an implicit equation for one field in terms of the other, it is easily solved by iteration. We therefore expect it to be the nonlinear approximation of choice for the comparison of large-scale structures and velocities in the quasi-linear regime, $\delta \sim 1 - 4$, where the effects of biased galaxy formation and the cosmological model might be separated.

This research was supported in part by NASA grant NAG-51228. We are grateful to A. Dekel, G. Ganon, and T. Kolatt for providing the $N$-body simulations used in §4 and for many useful discussions over the years. AY thanks the Yukawa Institute for Theoretical Physics of Kyoto University, its director, Professor Y. Nagaoka, and his host, Professor K. Tomita, for their kind hospitality during his visit there, when this work was performed.